\documentclass[conference]{IEEEtran}
\IEEEoverridecommandlockouts
\usepackage{cite}
\usepackage{amsmath,amssymb,amsfonts}
\usepackage{algorithmic}
\usepackage{graphicx}
\usepackage{textcomp}
\usepackage{xcolor}
\usepackage[utf8]{inputenc}
\usepackage[greek,english]{babel}
\usepackage{alphabeta}
\usepackage{soul}
\usepackage{subcaption}
\usepackage{float}
\usepackage[export]{adjustbox}
\usepackage{makecell}
\usepackage{xcolor}
\definecolor{cream}{RGB}{255,253,240}

\usepackage{fancyhdr}
\renewcommand{\baselinestretch}{0.95}
\newcommand{\freefootnote}[1]{%
  \begingroup
  \renewcommand{\thefootnote}{}
  \footnotetext{#1}
  \endgroup
}

\def\BibTeX{{\rm B\kern-.05em{\sc i\kern-.025em b}\kern-.08em
    T\kern-.1667em\lower.7ex\hbox{E}\kern-.125emX}}
\begin{document}

\title{On the Impact of Degradation-Balanced Scheduling in Manycore Systems}

\author{
\IEEEauthorblockN{Maria Pantazi-Kypraiou\textsuperscript{†},~George Goudroumanis\textsuperscript{†}, George~Floros\textsuperscript{‡}, Anuj Pathania\textsuperscript{¶}, and George Stamoulis\textsuperscript{†}}

\IEEEauthorblockA{
\textsuperscript{†}ECE Department, University of Thessaly, Volos, Greece, emails: \{mpantazi-, ggeorgios-r, georges\}@e-ce.uth.gr \\
\textsuperscript{‡} Department of Electronic and Electrical Engineering, Trinity College Dublin, Ireland, email: florosg@tcd.ie \\ 
\textsuperscript{¶}University of Amsterdam, The Netherlands, email: a.pathania@uva.nl
}}

% \makeatletter 
% \def\ps@IEEEtitlepagestyle{% 
% 	\def\@oddfoot{\mycopyrightnotice}% 
% 	\def\@evenfoot{}% 
% } 
% \def\mycopyrightnotice{% 
% 	{\footnotesize979-8-3503-1884-5/24/\$31.00 \textcopyright2024 IEEE\hfill}% <--- Change here 
% 	\gdef\mycopyrightnotice{}% just in case
% }

\maketitle
% \IEEEpubid{%
% \makebox[\columnwidth][l]{%
% \fontsize{8}{9}\selectfont
% 979-8-3195-1905-4/26/\$31.00~\textcopyright~2026 IEEE}
% \hspace{\columnsep}
% \makebox[\columnwidth]{}}
% \thispagestyle{icecsfirstpage}

\begin{abstract}
Electromigration degradation in manycore systems depends strongly on how runtime workload is distributed across cores. Task placement affects power, temperature, and current density, which in turn shape the spatial distribution of degradation over time. However, a scheduler that ranks cores only by their current reliability state may not always distinguish among candidate cores, especially when reliability values are close at decision time. In this case, scheduling decisions can be driven by secondary factors rather than by long-term degradation balance. This paper studies degradation-balanced scheduling and proposes a Reliability-Balanced (RB) policy that predicts the full-task effect of each candidate assignment. RB penalizes added imbalance in electromigration exposure, stress, active time, and task count while guarding the predicted Rvalue floor. On randomized continuous mixed workloads, the proposed scheduler preserves completed tasks while improving average Rvalue by 0.37\% and minimum Rvalue by 4.35\%. It reduces Rvalue, exposure, stress, and active-time standard deviation by 10.84\%, 9.39\%, 8.35\%, and 16.38\%, respectively.

\end{abstract}


\section{Introduction}

% Manycore processors improve throughput by executing many tasks in parallel, but their high power density and localized heating can accelerate long-term wear-out. Electromigration (EM) is one such reliability mechanism: current flow can move metal atoms in interconnects, producing voids, resistance increases, and eventual timing or connectivity failures \cite{TODO_EM_RELIABILITY}. EM degradation depends strongly on current density and temperature, both of which are affected by runtime task placement. As a result, a scheduler can influence not only immediate utilization and temperature, but also where degradation accumulates across the chip.

Manycore processors deliver high computational throughput through massive parallelism, but their increasing power density and thermal heterogeneity pose significant reliability challenges \cite{10.1145/3844139}. Among them, electromigration (EM) is a major degradation mechanism driven by current flow and accelerated by elevated temperature. Over time, EM can increase the resistance of the interconnect, create voids, and ultimately cause timing violations or connectivity failures \cite{bigalke2018need}. Consequently, reliability-aware scheduling has emerged as an important runtime mechanism to control how degradation accumulates across the chip \cite{kim2017energy}. 

\freefootnote{*This work was funded under the COIN-3D project, which has received funding from the European Union's Horizon Europe research and innovation program under grant agreement No. 101159667.}

EM reliability has traditionally been estimated using Black’s equation \cite{black2005electromigration}, which relates interconnect lifetime to current density and temperature. However, Black-based models provide aggregate lifetime estimates and do not explicitly capture the evolution of stress and degradation over time \cite{kahng2013potential}. To improve modeling fidelity, physics-based approaches derived from Korhonen’s stress evolution framework were introduced \cite{korhonen1993stress}, enabling the analysis of stress accumulation, void nucleation, and resistance growth under varying operating conditions \cite{kim2016cross, kim2017energy}. Subsequently, these models have been incorporated into system-level reliability management, task scheduling, and lifetime optimization frameworks for manycore and embedded systems \cite{kim2014lifetime}. However, existing schedulers generally assume that runtime reliability metrics provide sufficient decision-making value among candidate cores. Whether these metrics are sufficiently informative at decision time to meaningfully influence task placement remains largely unexplored.

Rather than relying solely on instantaneous reliability metrics, we examine the impact of estimating the reliability consequences of candidate task assignments and balancing degradation-related quantities across cores. The main contributions of this paper are threefold. \textit{First,} we investigate the impact of reliability-aware scheduling policies on degradation accumulation and distribution in manycore systems. \textit{Second,} we propose a Reliability-Balanced (RB) policy, a degradation-balancing scheduler that minimizes the imbalance in EM exposure, stress accumulation, active execution time, and task distribution while preserving a reliability guard. \textit{Finally,} we conduct an extensive experimental evaluation using continuous mixed workloads, demonstrating up to 16.4\% improvement in load balance and up to 10.8\% improvement in reliability balance while preserving task completion.

The remainder of this paper is organized as follows. Section \ref{sec:pf} presents the system model, scheduling environment, thermal model, and EM  formulation used in the paper. Section \ref{sec:scheduling} describes the evaluated scheduling policies, including the proposed RB policy. Finally, Section \ref{sec:exp} presents the evaluation results, while conclusions are drawn in Section \ref{sec:concl}.

\section{Problem Formulation} 
\label{sec:pf}

\subsection{Scheduling Environment}
We evaluate task placement in a discrete-time manycore scheduling environment with a fixed set of cores and a time-ordered workload of independent tasks. At each timestep, newly arrived tasks enter a visible waiting queue, following a standard parallel-machine scheduling abstraction ~\cite{1477600}. If the queue is non-empty, the scheduler selects one valid core for the first waiting task, and it is executed until its remaining duration reaches zero.

% The environment is intentionally work-conserving for the scheduling policies considered in this paper: tasks are not reordered by the environment after they enter the visible queue, and later tasks do not overtake earlier waiting tasks. This makes scheduling differences attributable to core-selection decisions rather than task-priority reordering. A core is considered valid for assignment only if it is idle, its Rvalue is above the configured reliability threshold, and its temperature is below the configured thermal safety threshold.
% The environment is intentionally work-conserving for the scheduling policies considered in this paper: tasks are not reordered by the environment after they enter the visible queue, and later tasks do not overtake earlier waiting tasks. This makes scheduling differences attributable to core-selection decisions rather than queue reordering. A core is considered valid for assignment only if it is idle, its reliability metric is above the configured reliability threshold, and its temperature is below the configured thermal safety threshold.

We simulate an abstract homogeneous 16-core manycore: each core executes at most one non-preemptive task, with task duration independent of the selected core. The model abstracts away caches, Network-on-Chip (NoC) routing, Dynamic Voltage and Frequency Scaling (DVFS), and floorplan coupling, and tracks scheduler-level interactions among task power, per-core temperature, reliability, exposure, stress, active time, and task count. A core is valid only if it is idle, its reliability metric is above the reliability threshold, and its temperature is below the thermal threshold. Since tasks do not overtake one another in the queue, policy differences come from core selection rather than task reordering.

\subsection{Task and Core Model}
Each task $\tau$ has an arrival time, an execution duration $d_{\tau}$, and a power demand $P_{\tau}$. 
% The selected policies dispatch the first visible waiting task, so the scheduling decision studied here is the choice of target core rather than task ordering.
Each core $i$ maintains an execution state and a reliability state,
\begin{equation}
  x_i = (T_i, R_i, E_i, \sigma_i, a_i, n_i),
\end{equation}
where $T_i$ is the temperature, $R_i$ is the current reliability score, $E_i$ is accumulated EM exposure, $\sigma_i$ is Korhonen-style stress, $a_i$ is the accumulated active time, and $n_i$ is the number of assigned tasks. The balance-oriented schedulers use these per-core distributions to avoid concentrating work and degradation on a small subset of cores.

\subsection{Thermal and Reliability Updates}
% \hl{ref + check first-order thermal update}
% \hl{begin-end equation!!!}

% % After assignment, active cores update their temperature and reliability state at every timestep while the task is running. The simulator uses a first-order RC-style thermal update,
% \[
%   T_i(t+1) =
%   T_i(t) + \alpha P_i(t) - \beta\left(T_i(t)-T_{\mathrm{amb}}\right),
% \]
% where idle cores use $P_i(t)=0$. In the current experimental configuration,
% $T_{\mathrm{amb}}=40^\circ$C, $\alpha=0.048$, and $\beta=0.08$.
% These parameters are simulator calibration constants: $T_{\mathrm{amb}}$ sets the ambient operating point, $\alpha$ controls heating per unit task power, and $\beta$ controls cooling toward ambient temperature. They are used to provide bounded thermal dynamics for comparing scheduling policies, rather than as technology-specific device parameters.

After assignment, active cores update their temperature and reliability state at every timestep while the task is running. Architecture-level thermal models commonly use compact RC representations to relate power dissipation to the transient temperature response~\cite{1206984,1650228}.
In single-node form, the thermal dynamics can be written as
\begin{equation}
  C_{\mathrm{th}}\frac{dT_i}{dt}
  =
  P_i(t)
  -
  \frac{T_i(t)-T_{\mathrm{amb}}}{R_{\mathrm{th}}},
\end{equation}
where $C_{\mathrm{th}}$ is the effective thermal capacitance and
$R_{\mathrm{th}}$ is the effective thermal resistance to ambient. Discretizing this first-order model gives the RC-style update used in the simulator,
\begin{equation}
  T_i(t+1) =
  T_i(t) + \alpha P_i(t) - \beta\left(T_i(t)-T_{\mathrm{amb}}\right),
\end{equation}
% where idle cores use $P_i(t)=0$. In the current experimental configuration, $T_{\mathrm{amb}}=40^\circ$C, $\alpha=0.048$, and $\beta=0.08$. These parameters are simulator calibration constants: $T_{\mathrm{amb}}$ sets the ambient operating point, $\alpha$ controls heating per unit task power, and $\beta$ controls cooling toward ambient temperature. They provide bounded thermal dynamics for comparing scheduling policies, rather than serving as technology-specific device parameters.
where idle cores use $P_i(t)=0$. We set $T_{\mathrm{amb}}=40^\circ$C as the nominal ambient point. The constants $\alpha=0.048$ and $\beta=0.08$ are scheduler-level calibration parameters: $\alpha/\beta$ sets the steady-state thermal rise, $T_{\mathrm{ss}}=T_{\mathrm{amb}}+(\alpha/\beta)P$, yielding about 70--112$^\circ$C for representative powers of 50--120, while $\beta$ sets a thermal time constant of about 12 timesteps.

EM degradation is updated after the thermal step by tracking a per-core stress state. The update is a reduced-order scheduler model rather than a full spatial solution of the Korhonen stress-diffusion equation. In one dimension, the Korhonen stress model can be written in the form:
\begin{equation}
    \frac{\partial \sigma}{\partial t}
    =
    \frac{\partial}{\partial x}
    \left[
    \frac{D(T)B\Omega}{k_BT_K}
    \left(
    \frac{\partial \sigma}{\partial x}
    + \frac{Z^* e \rho J}{\Omega}
    \right)
    \right],
\end{equation}

where $D(T)$ is the atomic diffusivity, $B$ is the effective modulus, $\Omega$ is the atomic volume, $T_K$ is the absolute temperature, $Z^*$ is the effective charge number, $e$ is the elementary charge, and $\rho$ is the resistivity. Our simulator does not represent individual wire geometries, boundary conditions, or spatial stress profiles; its reliability state is stored per core. We therefore do not solve this PDE directly. Instead, we approximate the spatial stress field with a single aggregate stress state $\sigma_i$ for each core and retain the part of the Korhonen model that determines how task assignment changes stress: the EM driving term scales with current density $J$ and temperature-dependent diffusivity $D(T)$. Absorbing geometry, material constants, and boundary effects into an effective coefficient gives the scheduler-level proportionality:

\begin{equation}
      \frac{d\sigma_i}{dt} \propto J_{\tau} D(T_i).
\end{equation}

The Arrhenius term is used because it is the temperature dependence of $D(T)$ in the Korhonen flux term. Atomic diffusion in metals is thermally activated, and EM treatments commonly model this diffusivity with an Arrhenius law~\cite{lienig2025fundamentals}:

\begin{equation}
    D(T_K)=D_0\exp\!\left(-\frac{E_a}{k_BT_K}\right).
\end{equation}

To express stress growth relative to a reference temperature, we normalize by $D(T_{\mathrm{ref},K})$:
\begin{equation}
    \begin{aligned}
    \frac{D(T_K)}{D(T_{\mathrm{ref},K})}
    &=
    \exp\!\left[
    -\frac{E_a}{k_B}
    \left(\frac{1}{T_K}-\frac{1}{T_{\mathrm{ref},K}}\right)
    \right] \\
    &\approx
    \exp\!\left(\gamma(T_i-T_{\mathrm{ref}})\right),
    \end{aligned}
\end{equation}

where the approximation uses $1/T_K \approx 1/T_{\mathrm{ref},K}-(T_K-T_{\mathrm{ref},K})/T_{\mathrm{ref},K}^2$, and $\gamma=E_a/(k_BT_{\mathrm{ref},K}^{2})$ is the local temperature sensitivity. Thus, the exponential factor in Eq. \ref{eq:final_stress} is not an independent heuristic; it is the normalized Arrhenius diffusivity from the Korhonen stress flux, expressed around $T_{\mathrm{ref}}$. Substituting this local approximation into the lumped stress law and absorbing constant factors into $c_{\sigma}$ gives, for a task with estimated current density $J_{\tau}$, the stress-rate law:

\begin{equation}
    \frac{d\sigma_i}{dt} = c_{\sigma} J_{\tau}
    \exp\!\left(\gamma(T_i(t)-T_{\mathrm{ref}})\right),
    \label{eq:final_stress}
\end{equation}

where $c_{\sigma}$ controls the effective stress growth rate and $T_{\mathrm{ref}}$ is the reference temperature. Applying a forward-Euler discretization,
$\sigma_i(t+\Delta t)=\sigma_i(t)+\dot{\sigma}_i(t)\Delta t$, gives the simulator update:
\begin{equation}
  \sigma_i(t+\Delta t) =
  \sigma_i(t) +
  c_{\sigma} J_{\tau}
  \exp\!\left(\gamma(T_i(t)-T_{\mathrm{ref}})\right)\Delta t,
\end{equation}
where the exponential term approximates the thermally accelerated EM stress growth relative to $T_{\mathrm{ref}}$. For positive $c_{\sigma}$, $J_{\tau}$, $\gamma$, and $\Delta t$, the increase is monotone in both current density and temperature:
\begin{equation}
  \frac{\partial \Delta\sigma_i}{\partial J_{\tau}} > 0,
  \qquad
  \frac{\partial \Delta\sigma_i}{\partial T_i} > 0.
\end{equation}
Thus, the reduced model is mathematically consistent with the scheduler-level assumption that hotter, higher-current assignments accumulate EM stress faster.

%%%%%%%%%%%%%%%%%%%%%%%%%%%%%%%%%%%%%%%%%%%%%%%%%%%%%%%%%%%%%%%%%%%%%%%%

% The simulator converts stress evolution into a bounded reliability indicator through normalized exposure. First, the current stress state gives a time-to-failure estimate $TTF_i$, defined as the time when $\sigma_i$ reaches the critical stress. This $TTF_i$ sets the lifetime scale:
The simulator converts stress evolution into a bounded reliability indicator through normalized exposure. First, the current stress state gives a time-to-failure estimate $TTF_i$, defined as the time when $\sigma_i$ reaches the critical stress. We use this deterministic stress-based estimate as the mean lifetime of the Weibull survival model used by the Rvalue metric~\cite{6974677}. Since a Weibull lifetime with scale $A_i$ and shape $\beta_w$ has mean $A_i\Gamma(1+1/\beta_w)$, matching this mean to $TTF_i$ gives:

\begin{equation}
    A_i =
  \frac{TTF_i}{\Gamma\!\left(1+\frac{1}{\beta_w}\right)}.
\end{equation}

Normalized exposure then accumulates as
\begin{equation}
  E_i(t+\Delta t) = E_i(t) + \frac{\Delta t}{A_i}.
\end{equation}
The Rvalue is calculated from this accumulated exposure as:
\begin{equation}
    R_i(t) =
  \exp\!\left(-E_i(t)^{\beta_w}\right),
\end{equation}

where $\beta_w$ controls the shape of the reliability curve. Therefore $R_i \in [0,1]$: it starts at 1 for an undegraded core and decreases as exposure accumulates. The scheduler uses this non-increasing Rvalue both for validity checks and for final reliability evaluation.

\section{Scheduling Policies}
\label{sec:scheduling}
% We present the scheduling policies as an evolution from simple runtime reliability awareness to predictive degradation balancing. Round Robin is kept only as a simple reference baseline: it ignores temperature, ARB, exposure, and stress during ranking, and cycles through core indices until the environment accepts an idle valid core.

\subsubsection{Rvalue-Temperature Policy}
The Rvalue-temperature scheduler selects the valid core with the largest current-state score,
\[
  s_i^{R-T} = R_i - \lambda T_i,
\]
where $\lambda=0.01$ is a heuristic temperature-penalty weight. This policy captures a common rule-based intuition: prefer reliable cores while discouraging assignments to thermally stressed cores. Its limitation is that it relies only on the current reliability state. When candidate Rvalues are compressed or tied, temperature and tie-breaking behavior can dominate the decision, even though the policy is nominally reliability-aware. The policy scans the cores to identify valid candidates. It evaluates a scalar score per candidate, giving $O(N)$ worst-case selection cost, or $O(C)$ once the $C$ valid candidates are available.

\subsubsection{Horizon Exposure Policy}
Horizon Exposure is a predictive scheduler that evaluates future exposure growth before selecting a core.
% For each valid candidate core $c$, the scheduler copies the current simulator state, assigns the waiting task to $c$, and then rolls out up to $K$ subsequent task assignments using a fixed greedy rollout policy.
For each valid candidate core $c$, the scheduler copies the current simulator state, assigns the waiting task to $c$, and then rolls out up to $K=8$ subsequent task assignments. The rollout uses the same greedy minimum-exposure-growth rule at each simulated decision. The simulated time covered by this rollout is denoted by $H$, which depends on the durations and arrival times of the rolled-out tasks. If $E_i^{0}$ is the current exposure of the core $i$ and $E_i^{K}(c)$ is the predicted exposure after the rollout, the candidate cost is:
\[
  G_E(c)=\sum_i \max\left(0, E_i^{K}(c)-E_i^{0}\right).
\]
The selected core is the candidate with the smallest predicted cumulative exposure growth. This policy tests whether looking beyond the current Rvalue can provide a more informative reliability signal. However, exposure growth is a global consumption measure; by itself, it does not explicitly penalize uneven exposure, stress, or workload distribution between cores. If $H$ is the total number of simulated timesteps in the rollout, the simulator cost is $O(C H N)$ because each candidate rollout updates all cores at each simulated timestep.

\subsubsection{Proposed Reliability-Balanced (RB) Policy}
% The final proposed policy is designed for reliability-aware load and degradation balancing. Its objective is to avoid degradation hotspots by minimizing the incremental imbalance caused by assigning the current task to each core.
The proposed RB policy treats reliability-aware scheduling as a degradation balancing problem. For each candidate core, it predicts the post-task state and penalizes assignments that create exposure, stress, or workload hotspots. The predicted post-assignment state is used to evaluate how evenly degradation would remain distributed across the chip. Candidate assignments that increase exposure, stress, or workload imbalance receive lower scores.

% For each candidate core, the scheduler predicts the reliability impact of assigning the incoming task to that core. The scheduler then uses the predicted post-assignment state to evaluate how evenly degradation remains distributed across the chip. Candidate assignments that increase exposure, stress, or workload imbalance receive lower scores.

\[
  \Delta_E(c) =
    \operatorname{std}(\mathbf{E}^{after}(c)) -
    \operatorname{std}(\mathbf{E}^{before}),
\]
with analogous terms $\Delta_{\sigma}(c)$, $\Delta_a(c)$, and $\Delta_n(c)$
for stress, active time, and task-count imbalance.

RB selects the candidate with the highest score
\[
\begin{aligned}
  s_c ={}&
  - w_E \Delta_E(c)
  - w_{\sigma} \Delta_{\sigma}(c)
  - w_a \Delta_a(c)
  - w_n \Delta_n(c) \\
  &+ w_{R}\min_i R_i^{after}(c).
\end{aligned}
\]
% The weights used in the reported experiments are $w_E=1.0$, $w_{\sigma}=0.5$, $w_a=0.5$, $w_n=0.25$, and $w_{R}=0.25$. The first four terms penalize assignments that increase degradation or load imbalance, while the final term acts as a guard against assignments that would reduce the predicted system Rvalue floor. The score therefore favors assignments that keep the predicted post-task state more even across cores while avoiding a large drop in the system Rvalue. Its effect is evaluated through distributional metrics such as exposure, stress, Rvalue, and active-time dispersion, together with completed-task count. For a task with remaining duration $d_{\tau}$, RB simulates the full current task for each candidate but does not roll out additional future tasks. Its simulator cost is therefore $O(C d_{\tau} N)$, making it more expensive than the myopic policies but avoiding the extra horizon length of Horizon Exposure.

% The weights used in the reported experiments are $w_E=1.0$, $w_{\sigma}=0.5$, $w_a=0.5$, $w_n=0.25$, and $w_{R}=0.25$. The imbalance terms penalize added spread in degradation and load, whereas the Rvalue term protects against assignments that reduce the predicted system reliability floor. RB simulates the full current task for each candidate, but does not roll out additional future tasks. For a task with remaining duration $d_{\tau}$, its simulator cost is $O(C d_{\tau} N)$, making it more expensive than the myopic policies but avoiding the extra horizon length of Horizon Exposure.

The reported experiments use one fixed weight vector: $w_E=1.0$, $w_{\sigma}=0.5$, $w_a=0.5$, $w_n=0.25$, and $w_{R}=0.25$. Here, $w_E$ sets exposure as the reference balance term, while $w_{\sigma}$, $w_a$, and $w_n$ penalize stress, active-time, and task-count imbalance. The $w_R$ term guards the predicted minimum Rvalue. All weights are fixed across reported runs and are not retuned per seed. We selected them from a preliminary sensitivity screening comparing pure-balance, guarded, aggressive-balance, and load-heavy variants. RB predicts only the current task, giving $O(C d_{\tau} N)$ cost: higher than myopic policies, but lower than rollout-based Horizon Exposure.
\section{Experimental Evaluation}
\label{sec:exp}
% All experiments in the main evaluation use a 16-core simulated manycore system with the corrected Korhonen-inspired reliability model and the RC-style thermal model described in Section~II. The main workload is \texttt{randomized\_continuous\_mixed\_reliability}, which produces continuous task arrivals with mixed duration and power demand. Results are averaged over 20 random seeds. We compare four schedulers: Round Robin, the Rvalue-temperature rule, the Horizon Exposure policy, and the final proposed RB policy. Because all
% policies complete the same number of tasks on average, the comparison primarily
% reflects how each policy distributes work and degradation, rather than a
% throughput difference.

The main evaluation uses the abstract 16-core manycore platform described in
Section \ref{sec:pf}. We evaluated a set of 20 randomized continuous-arrival workloads with overlapping tasks and mixed duration and power demand. We compare Round Robin, the Rvalue-temperature rule, Horizon Exposure, and the proposed RB policy. Because all policies complete the same number of tasks on average, the comparison mainly reflects how each policy distributes work and degradation, rather than a throughput difference.

\subsection{Limitations of Current-Rvalue Scheduling}
Before evaluating the proposed scheduler, we first examine whether the reliability state is sufficiently visible at the scheduling time. In the decision-time diagnostic, candidate Rvalues were often compressed, that is, the mean Rvalue range among valid candidates was $5.43\times 10^{-6}$, the median range was zero, and 96.9\% of the decisions were effectively tied in Rvalue. Under the same score, temperature dominated candidate separation in 56.2\% of decisions. This suggests that a myopic current-Rvalue score can behave more like a temperature-aware selector than a degradation-aware scheduler when candidate Rvalues are nearly identical.

% Additional candidate-availability diagnostics show the same trend from another angle. In multi-candidate decisions, the median best-versus-second-best Rvalue gap was zero, while the median TTF gap was nonzero. Thus, TTF can be more discriminative than the Rvalue at decision time. However, later correlation diagnostics showed that local TTF-related signals were weak predictors of final system reliability. These observations motivate a balance-aware objective: instead of optimizing only the weakest predicted TTF, the scheduler should reduce concentration of exposure, stress, and activity across cores.

These observations motivate a balance-aware objective: instead of relying only on the current Rvalue of candidate cores, the scheduler should reduce the concentration of exposure, stress, and activity across the core population.

% \subsection{Continuous Mixed Reliability Workload}
% The main workload is designed to avoid a purely burst-dominated setting. Tasks arrive throughout the episode, overlap in execution, and include a mixture of moderate and high power levels, creating repeated scheduling opportunities. Because all policies complete the same number of tasks on average, the comparison primarily reflects how each policy distributes work and degradation, rather than a throughput difference. 
% The representative timeline in Fig.~\ref{fig:continuous_mixed_timeline} illustrates the task-placement patterns observed for seed 1.

% \begin{figure}[t]
% \centering
% \includegraphics[width=\columnwidth]{figures/task_assignment_timeline_seed1.png}
% \caption{Representative seed-1 task assignment timeline for the continuous
% mixed workload. This figure compares how Round Robin, Current ARB-temp,
% Horizon Exposure, and ReliabilityBalancedGuarded distribute task assignments
% over time.}
% \label{fig:continuous_mixed_timeline}
% \end{figure}

\begin{figure}[t]
\centering
\includegraphics[width=\columnwidth]{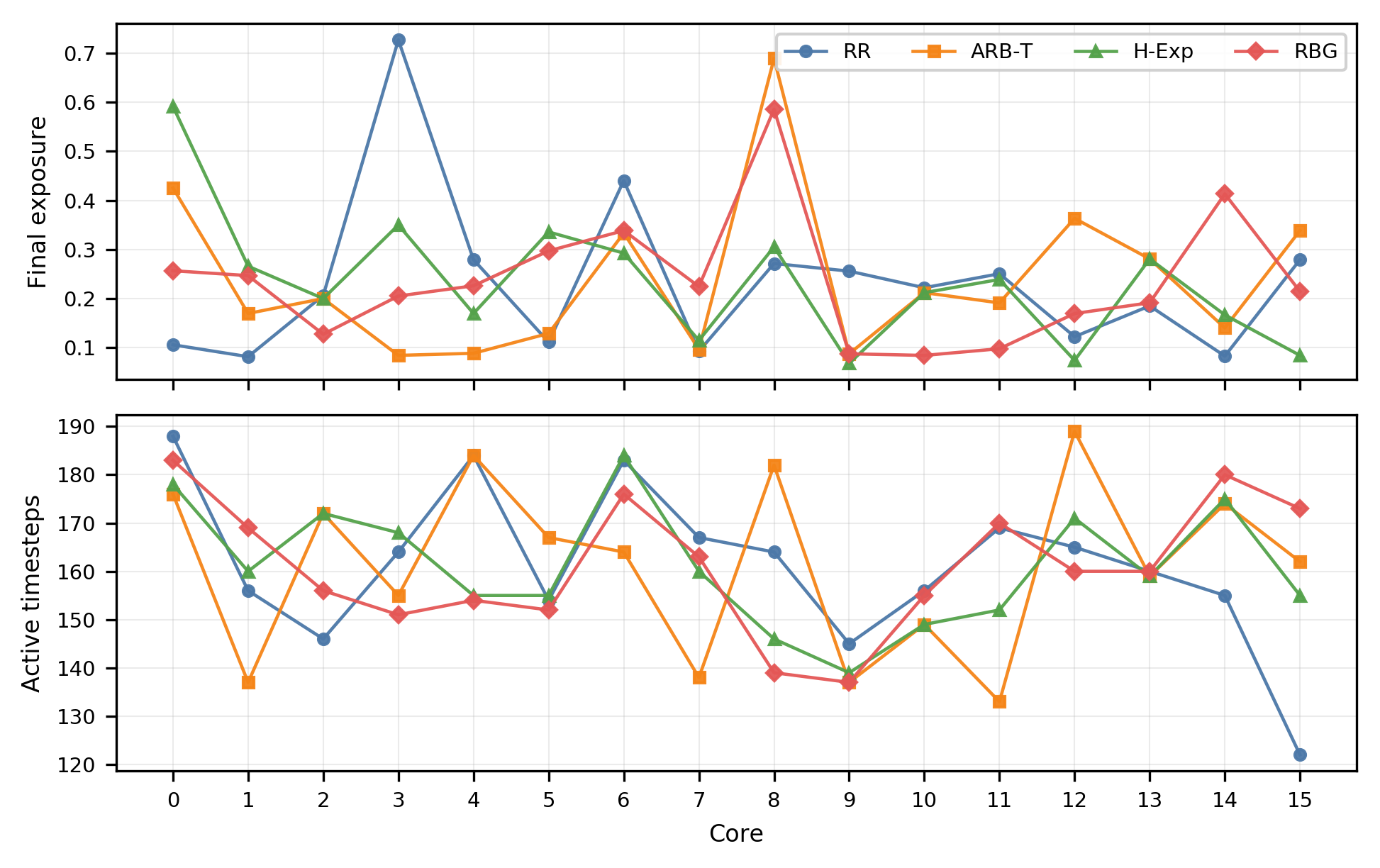}
\caption{Per-core degradation and load balance. The upper panel shows final EM exposure, and the lower panel shows active timesteps. Lower dispersion indicates less concentration of degradation and work across cores.}
\label{fig:continuous_mixed_balance}
\end{figure}

% \begin{table}[t]
% \centering
% \caption{Relative improvement over Current Rvalue-temp on the continuous mixed workload. Positive values are better; for standard-deviation metrics, positive values indicate reduced imbalance.}
% \label{tab:continuous_mixed_main}
% \scriptsize
% \setlength{\tabcolsep}{3.5pt}
% \begin{tabular}{lrrr}
% \hline
% Scheduler & Min Rvalue & Rvalue std. & Deg. diff. \\
% \hline
% Round Robin & -5.6\% & -7.8\% & -12.3\% \\
% Current Rvalue-temp & 0.0\% & 0.0\% & 0.0\% \\
% Horizon Exposure & +1.5\% & +5.7\% & +3.5\% \\
% \textbf{RB} &
% \textbf{+4.4\%} & \textbf{+10.8\%} & \textbf{+10.3\%} \\
% \hline
% \end{tabular}

% \vspace{0.6mm}

% \begin{tabular}{lrrr}
% \hline
% Scheduler & Exp. std. & Stress std. & Act. std. \\
% \hline
% Round Robin & -5.9\% & -4.4\% & -5.1\% \\
% Current Rvalue-temp & 0.0\% & 0.0\% & 0.0\% \\
% Horizon Exposure & +6.0\% & +6.5\% & +2.4\% \\
% \textbf{RB} &
% \textbf{+9.4\%} & \textbf{+8.3\%} & \textbf{+16.4\%} \\
% \hline
% \end{tabular}
% \end{table}

\begin{table*}[!hbt]
\centering
\caption{Relative improvement over current Rvalue-temp on the continuous mixed workload. Positive values are better. For standard-deviation metrics, positive values indicate reduced imbalance.}
\label{tab:continuous_mixed_main}
% \footnotesize
\setlength{\tabcolsep}{12pt}
\begin{tabular}{lrrrrrr}
\hline
Scheduler & Min Rvalue & Rvalue std. & Deg. diff. & Exp. std. & Stress std. & Act. std. \\
\hline
Round Robin & -5.6\% & -7.8\% & -12.3\% & -5.9\% & -4.4\% & -5.1\% \\
% Current Rvalue-temp & 0.0\% & 0.0\% & 0.0\% & 0.0\% & 0.0\% & 0.0\% \\
Horizon Exposure & +1.5\% & +5.7\% & +3.5\% & +6.0\% & +6.5\% & +2.4\% \\
\textbf{RB} &
\textbf{+4.4\%} & \textbf{+10.8\%} & \textbf{+10.3\%} &
\textbf{+9.4\%} & \textbf{+8.3\%} & \textbf{+16.4\%} \\
\hline
\end{tabular}
\end{table*}

\subsection{Scheduler Comparison}
% Table~\ref{tab:continuous_mixed_main} summarizes the main evaluation. Round Robin is included as a simple non-reliability-aware reference. The current Rvalue-temperature scheduler is the myopic reliability-aware starting point, and Horizon Exposure is the intermediate predictive policy. The table reports relative improvement over the current Rvalue-temperature scheduler, using the latter as the zero-reference baseline. Positive values are better. For standard-deviation metrics, positive improvement means reduced dispersion. The current Rvalue-temperature baseline achieves an average Rvalue of 0.922, a minimum of 0.684, Rvalue standard deviation of 0.090, exposure standard deviation of 0.156, stress standard deviation of 40.401, and active-time fairness of 0.989.

Table~\ref{tab:continuous_mixed_main} summarizes the main evaluation. The results are reported as a relative improvement over the Rvalue-temperature scheduler, which is used as the zero-reference baseline. Positive values are better; for standard-deviation metrics, they indicate reduced dispersion. The baseline achieves mean and minimum Rvalues of 0.922 and 0.684, respectively, with Rvalue, exposure, and stress standard deviations of 0.090, 0.156, and 40.401.

% Compared with Current Rvalue-temp, \textbf{RB} improves minimum Rvalue by 4.4\%, reduces Rvalue standard deviation by 10.8\%, reduces maximum degradation difference by 10.3\%, reduces exposure standard deviation by 9.4\%, reduces stress standard deviation by 8.3\%, and reduces active-time standard deviation by 16.4\%, while preserving the same number of completed tasks. These trends support the intended contribution: the proposed policy reduces degradation concentration while maintaining task completion. The per-core exposure and active-time distributions in Fig.~\ref{fig:continuous_mixed_balance} show the same effect at the core level for the representative workload instance.

Horizon Exposure improves several Rvalue and exposure metrics, showing that predictive reliability information can help. However, exposure growth alone does not explicitly optimize degradation balance, and the multi-step rollout is substantially more expensive in the current implementation.

% Horizon Exposure improves several Rvalue and exposure metrics, confirming that future reliability prediction is useful. However, exposure growth alone does not explicitly optimize degradation balance, and the multi-step rollout is substantially more expensive in the current implementation. \textbf{RB} is the final proposed scheduler: it keeps prediction local to the current task but directly optimizes the imbalance terms that define the paper's degradation-balancing objective.

Compared with current Rvalue-temp, \textbf{RB} improves minimum Rvalue by 4.4\%, reduces Rvalue standard deviation by 10.8\%, reduces maximum degradation difference by 10.3\%, reduces exposure standard deviation by 9.4\%, reduces stress standard deviation by 8.3\%, and reduces active-time standard deviation by 16.4\%, while preserving the same number of completed tasks. The per-core exposure and active-time distributions in Fig.~\ref{fig:continuous_mixed_balance} show the same effect at the core level: RB reduces degradation concentration by directly optimizing imbalance terms rather than only cumulative exposure growth.

\subsection{Discussion}

The results should be interpreted as degradation-balance improvements rather than as a claim that one reliability metric alone is sufficient for scheduling. They also show a cost tradeoff among the policies. Round Robin and Rvalue-temp have similar measured runtime because both are myopic policies. Horizon Exposure is the most expensive due to rollout simulation, while RB is lower cost because it predicts only the current task. In our implementation, RB was about $14\times$ faster than Horizon Exposure, although both predictive methods were slower than the myopic baselines.

RB is evaluated as a simulator-level placement heuristic invoked at task
dispatch, not as an OS or firmware implementation. All policies complete the
same number of tasks in our workload, while real-system scheduling latency and
integration overhead remain future work.

% An important observation is that degradation balance and worst-core TTF are different objectives. Reducing exposure and stress concentration does not necessarily imply a consistent improvement in worst-core or fifth-percentile TTF. This is a limitation of the current formulation rather than a contradiction of the proposed objective. The evaluation is also limited to a custom scheduling environment and synthetic workloads. Validation with more detailed thermal, architectural, and interconnect simulation frameworks is left for future work.
\section{Conclusion}
\label{sec:concl}
% \hl{This paper studied reliability-aware task scheduling under a Korhonen electromigration model, with the goal of reducing degradation hotspots rather than maximizing a single worst-core lifetime metric. Starting from a myopic ARB-temperature scheduler, we observed that current ARB values can be weakly discriminative at decision time. We then evaluated predictive exposure-aware scheduling and proposed ReliabilityBalancedGuarded, which directly minimizes incremental degradation and load imbalance while guarding the ARB floor. Across the continuous mixed workload, the proposed scheduler improves ARB, exposure, stress, and active-time balance while preserving task completion. Future work will validate the approach with richer thermal and architectural models and study broader workload classes.}

This paper evaluated the impact of reliability-aware scheduling policies on EM degradation in manycore systems. More specifically, we compared scheduling approaches ranging from current-state reliability metrics to predictive reliability estimation and proposed RB, a scheduler that balances degradation-related quantities while preserving a reliability guard. Experimental evaluation in continuous randomized mixed workloads demonstrated that the proposed approach improves degradation distribution and system reliability, increasing the minimum Rvalue by 4.4\% and reducing Rvalue, exposure, stress, and active-time imbalance by up to 10.8\%, 9.4\%, 8.3\%, and 16.4\%, respectively, without affecting task completion.

% \section*{Acknowledgment}
% This work was funded under the COIN-3D project, which has received funding from the European Union’s Horizon Europe research and innovation program under grant agreement No. 101159667. The content reflects only the authors’ views, and the European Commission is not responsible for any use that may be made of the information it contains.

\bibliographystyle{IEEEtran}
\bibliography{bibliography}

\end{document}